\documentclass[a4paper]{jpconf}

\newcommand{\nin}{\noindent}
\newcommand{\be}{\begin{equation}}
\newcommand{\ee}{\end{equation}}
\newcommand{\bea}{\begin{eqnarray}}
\newcommand{\eea}{\end{eqnarray}}
\newcommand{\nn}{\nonumber\\}

\usepackage{graphicx}

\begin{document}
\title{Non-extensive statistics and its effects on cosmology}

\author{Ariadne Vergou}

\address{King's College London, Dept. of Physics, Strand, London, WC2R 2LS, UK}

\ead{ariadni.vergou@kcl.ac.uk}

\begin{abstract}
We apply non-extensive statistics, namely Tsallis statistics, on a case of supercritical string cosmology (SSC) as the one studied in \cite{liouville} and obtain interesting cosmological modifications. At the beginning, we seek for the non-extensive corrections to the dilaton energy density and the off-shell (off-critical) terms and by using the set of dynamical equations of \cite{liouville}, we derive the modified evolution equation for the radiation energy density in a r.d.e.(radiation dominated era). This in fact is characterised by fractal( "exotic") scaling and this seems to be a generic result of our analysis. The modified Boltzmann equation was also considered , giving us the effects on dark matter candidates relic abundances of  non-extensitivity besides the SSC effects. An effort to give a physical interpretation to the essential results of models such as the Tsallis statistical model has been done in collaboration with N. Mavromatos and S.Sarkar by using the D-particles foam model which is presented at the last part of this work.

\end{abstract}

\section{Introduction:Tsallis statistics framework}
Tsallis statistics formalism is based on considering non-extensive (non-additive) entropies, which can reproduce the standard Boltzmann-Gibbs entropy at a certain limit of the free parameter of the theory,known as the non-extensive parameter, often denoted as q.The important thing about these families of entropies is that, when applied to ordinary statistical mechanics, they give rise to probabilities that follow power laws instead of the exponential laws of the standard case( for details on this see \cite{non-ext}). In fact this last property makes Tsallis approach a very convenient tool in the attempt to describe systems with long range interactions, e.g gravitation, or systems having long memory effects. In most cases that Tsallis formalism is adopted ,e.g.\cite{pessah}, the non-extensive parameter is taken to be constant and very close to the value for which ordinary statistical mechanics is obtained ($q=1$), meaning that all calculations can be performed in the leading order to ($q-1$), with no considerable loss of information. These two assumptions are also made throughout our work whereas in general it would be interesting to consider,at some point, a time-dependent parameter q, with a function dependancy that could be treated as a fitting parameter of our model. 

It is also important to mention that although Tsallis entropy gives probabilities which have physically, a different behaviour than in the standard case, it is constructed in such a way that it preserves all the "good" properties of the Boltzmann-Gibbs entropy. Namely, it is always positive, it is concave which is crucial for thermodynamical stability and it also preserves the Legendre transform structure of thermodynamics.This in fact is what differentiates Tsallis approach with other attempts made to walk away from extensitivity.

\section{Tsallis statistics effects on SSC: a case study}
We consider the SSC model adopted in \cite{liouville} and seek for the modifications due to non-extensitivity. The set of dynamical equations in the Einstein frame in this case, will have the form:
\be\label{one}
3\hat{H}^2=\tilde{\rho}_m+\tilde{\rho}_\phi+\frac{e^{2\phi}}{2}\tilde{G}_\phi
\ee

\be\label{two}
2\dot{\hat{H}}=-\tilde{\rho}_m-\tilde{\rho}_\phi-\tilde{p}_m-\tilde{p}_\phi+\frac{\tilde{G}_{ii}}{a^2}
\ee

\be\label{three}
\ddot{\phi}+3\hat{H}\dot{\phi}+\frac{1}{4}\frac{\vartheta\hat{V}_{all}}{\vartheta\phi}+\frac{1}{2}(\tilde{\rho}_m-3\tilde{p}_m)=-\frac{3}{2}\frac{\tilde{G}_{ii}}{a^2}-\frac{e^{2\phi}}{2}\tilde{G}_\phi
\ee

\nin where all densities now carry the non-extensive corrections. We also make clear that in the non-critical case, the central charge deficit that comes into the potential for the dilaton field is not constant but evolves with time through the Curci-Paffuti equation (for details see \cite{liouville} and \cite{lahanas}).
The equilibrium energy densities for matter (non-relativistic limit) and radiation are given in \cite{pessah}.By using these expressions, as well as a correction of the form $e^{\int \Gamma dt}$ where $\Gamma$ accounts for the off-shell, dilaton terms and can be overall seen  as a source term, we can explicity write the contributions of the ordinary matter to the energy density $\tilde{\rho}_m$ appearing in (\ref{one}),(\ref{two}),(\ref{three}).

For instance the non-relativistic energy density will be given by:
\be
\rho_{b,q}(t)=\frac{\alpha(t_0)^3}{\alpha(t)^3}g_bm(\frac{mT_0}{2\pi})^\frac{3}{2}e^{-(m-\mu)/T_0}[1+\frac{q-1}{2}(\frac{15}{4}+3\frac{m-\mu}{T_0}+(\frac{m-\mu}{T_0})^2)]e^{\int\limits_{t_0}^{t} \Gamma dt}
\ee
\nin where it is easy to recognize the standard expression and the two corrections.

In equations (\ref{one}), (\ref{two}) and (\ref{three}) we have neglected any non-extensive corrections to the off-critical terms $\tilde{G}_\phi$ and $\tilde{G}_{ii}$, which is a valid approximation under the assumption that those terms are of order less than ($q-1$).Therefore all that we are left with and always to leading order in ($q-1$) is to determine the non-extensive corrections to the dilaton energy density and the exotic matter which is also taken into account in the density $\tilde{\rho}_m$ by the authors of \cite{liouville}. By assuming a radiation dominated era and by introducing a convenient effective number of degrees of freedom defined by the relation:
\be
\tilde{g}_{eff,q}=g_{eff,q}+\frac{30}{\pi^2}T^{-4}\Delta\rho
\ee
\nin where $\Delta\rho$ incorporates all the dilaton and off-shell terms, which are assumed not to be thermalized and by using for $g_{eff,q}$ the expression given in \cite{pessah}:
\be
g_{eff,q}=\sum g_{i,bosons}(\frac{T_i}{T})^4+\frac{7}{8}\sum g_{j,fermions}(\frac{T_j}{T})^4+(q-1)[9.58\sum g_{i,bosons}(\frac{T_i}{T})^4+8.98\sum g_{j,fermions}(\frac{T_j}{T})^4]
\ee
\nin we can obtain the correction to the dilaton field energy density:
\be
\tilde{\rho}_{\phi,q}=\dot{\phi}^2+\frac{\hat{V}_{all}}{2}+\frac{\pi^2}{30}\frac{0.3^2}{g_{eff,q}}\frac{\tilde{\rho}_r}{3\hat{H}^2}\frac{1}{t_E^2}(q-1)[9.197\sum g_{bosons} +8.623\sum g_{fermions}]
\ee

For the exotic matter we considered that any q-dependance will come into its equation of state parameter that anyway is treated as a fitting parameter in the numerical analysis performed to (\ref{one}),(\ref{two}),(\ref{three})( see also \cite{lahanas}).

With all these in hands we can write down the modified continuity equations stemming from combination of (\ref{one}),(\ref{two}) and (\ref{three}):

\be\label{cont}
\frac{d\tilde{\rho}_m}{dt_E}+3\hat{H}(\tilde{\rho}_m+\tilde{p}_m)-\dot{\phi}(\tilde{\rho}_m-3\tilde{p}_m)+6\hat{H}C+\frac{dC}{dt_E}=6(\hat{H}+\dot{\phi})\frac{\tilde{G}_{ii}}{a^2}
\ee
\nin where by C we have denoted the q-correction to the dilaton energy density,i.e:
\be
C=\frac{\pi^2}{30}\frac{0.3^2}{g_{eff,q}}\frac{\tilde{\rho}_r}{3\hat{H}^2}\frac{1}{t_E^2}(q-1)[9.197\sum g_{bosons} +8.623\sum g_{fermions}]
\ee

Now from (\ref{cont}) we can easily obtain the evolution equation for radiation which will have the form:
\be\label{rad}
\frac{d\tilde{\rho}_r}{dt_E}(1+\frac{\epsilon}{3\hat{H}^2t_E})=-4\hat{H}\tilde{\rho}_r-\frac{2\epsilon}{\hat{H}t_E^2}[1-\frac{1}{3\hat{H}t_E}-\frac{1}{3}\frac{\hat{\dot{H}}}{\hat{H}^2}]\tilde{\rho}_r
\ee
\nin where the parameter $\epsilon$ appearing above is defined through: 
\be
\epsilon\equiv\frac{3\hat{H}^2t_E^2}{\tilde{\rho}_r}\times C
\ee

One can try to solve (\ref{rad}) perturbatively in $\epsilon$ since it contains ($q-1$) which can always be made arbitrarily small by chosing appropriately the non-extensive parameter q. We obtain:

\be
\tilde{\rho}_r(t_E)\approx [\tilde{\rho}_{r,in}\alpha^{-4}]e^{-\frac{2}{3}\epsilon ln t_E}=\tilde{\rho}_{r,in}\alpha^{-4-\delta}
\ee
\nin where we have set:
\be
\delta\equiv\frac{2}{3}\epsilon\frac{ln t_E}{ln\alpha}=\frac{4}{3}\epsilon
\ee

From (\ref{rad}) and the definition for $\delta$ it is clear that under the assumption of non-extensitivity and for the case that we are studying, radiation scales in a fractal way and this result arises naturally in our analysis.

\begin{figure}[th]
\centering
\includegraphics[width=12.6cm, height=8.4cm]{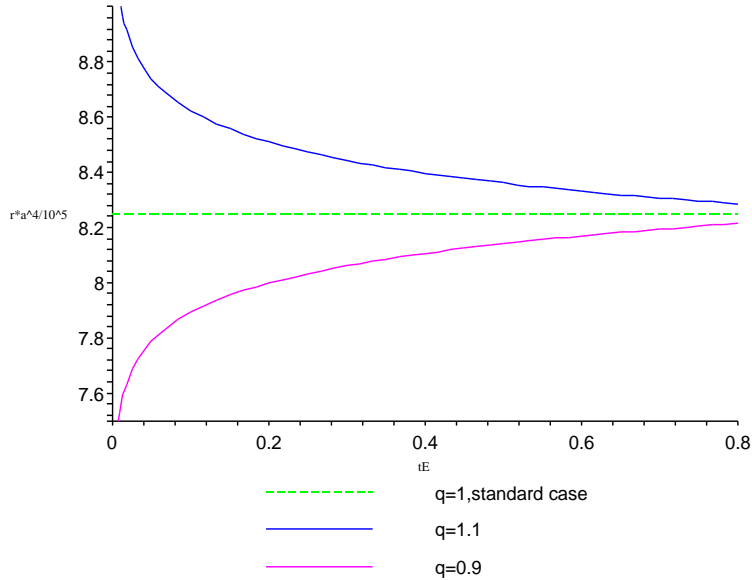}
\caption{{\footnotesize: Plot of the radiation energy density $\times \alpha^4$ with respect to the Einstein time}}
\label{fig1}
\end{figure}

In the above plots we have used the numerical estimation $\epsilon\approx0.2885(q-1)$ calculated for a r.d.e.

\subsection{Non-extensive effects on Boltzmann equation}
In all the following analysis, we restrict our attention to supersymmetric dark matter species, such as neutralinos, which are thaught to be the lightest supersymmetric particles (LSP). In fact such an analysis can be related to a quite interesting and rich phenomenology (see also \cite{boltzman}) and as such it can be a very promising tool. In order to describe the number density evolution of those species, one needs to write down the modified Boltzmann equation due to non-extensitivity and the off-shell/dilaton terms (we recall that our case study is a SSC case). Those last corrections for the standard case, are given in \cite{boltzman}.

The corrected Boltzmann equation for a particular species $\psi$  of mass $m_\psi$ and in terms of the parameters $Y_\psi\equiv\frac{n_\psi}{s}$ (number per entropy density) and $\chi\equiv\frac{T}{m}$ has the form:
\be\label{boltz}
\frac{dY_{\psi,q}}{d\chi}=0.264\left\langle \sigma_A\left|\upsilon\right|\right\rangle\tilde{g}_{eff,q}^{-\frac{1}{2}}h_q ~m~m_{pl}[(Y_{\psi,q})^2-(Y_{\psi,q,eq})^2]-\frac{\Gamma H^{-1}}{\chi}Y_{\psi,q}
\ee
\nin where $h_q$ denotes the entropy degrees of freedom and its form in the non-extensive case is given in \cite{pessah}.

By looking at (\ref{boltz}) we see that this has exactly the same form as in \cite{boltzman} with the only difference that the number densities as well as the effective number of degrees of freedom and the entropy degrees of freedom all carry the non-extensive corrections. We note here that the last term in (\ref{boltz}) represents the contributions from the extra terms (off-shell and dilaton) which are elegantly incorporated in the $\Gamma$ term.

Following the standard approach in solving the Boltzman equation, let us assume that the species under consideration decouples from the plasma at a specific temperature $\chi_f$ (freeze-out point). It is clear that at temperatures greater than $\chi_f$, the interactions of the species maintain it in thermal equilibrium, therefore  $Y_{\psi,q}=Y_{\psi,q,eq}$ at all those times.Substituting this in (\ref{boltz}), yields:
\be\label{equil}
Y_{\psi,q}(\chi)=0.145g_s(h_q)^{-1}\chi^{-\frac{3}{2}}e^{-\frac{1}{\chi}}[1+\frac{q-1}{2}(\frac{15}{4}+3\chi^{-1}+\chi^{-2})]e^{-\int\limits_{\chi_{in}}^{\chi}\frac{\Gamma H^{-1}}{\chi}}
\ee
\nin and $g_s$ counts the particle's spin degrees of freedom.

In order to determine the freeze-out point, we use the general freeze-out criterion: $\Gamma(\chi_f)\approx H(\chi_f)$ and we also evaluate \ref{equil} at $\chi_f$. Therefore we obtain the "new" freeze-out point that will be given by:
\be  
\chi_{f,q}=\chi_f-\frac{q-1}{2}[\frac{1}{2}\chi_f^{-1}+\chi_f^{-2}][\frac{15}{4}+3\chi_f^{-1}+\chi_f^{-2}-\frac{1}{\tilde{g}_{eff,q}}(9.58\sum g_{i,bosons} +8.98\sum g_{j,fermions})]
\ee
\nin where all effective number of degress of freedom are counted at the freeze-out point.
We also note that by $\chi_f$ in (\ref{equil}) we mean the "standard" freeze-out point satisfying the relation (\cite{boltzman}):
\be\label{freez}
\chi_f^{-1}=ln[0.038\left\langle \sigma_A\left|\upsilon\right|\right\rangle_{\chi_f}g_{eff}^{-\frac{1}{2}}\chi_f^{-\frac{1}{2}}~m~m_{pl} g_s]+\frac{1}{2}ln(\frac{g_eff}{\tilde{g_{eff}}})+ \int\limits_{\chi_{f}}^{\chi_{in}}\frac{\Gamma H^{-1}}{\chi}
\ee

From (\ref{freez}) we see that the non-extensive correction to the freeze-out point of the species $\psi$ can be either positive or negative depending on how early or late the species decoupled from the plasma. Very roughly we could say that if the species decoupled at early times (large $\chi_f$) where also the relativistic contributions were quite large, then with the non-extensitivity taken into account the species should have decoupled even earlier, whereas for small freeze-out points (late decoupling) the species in the non-extensive case should have decoupled even later.

With all these in hand and by following exactly the same approximations as in \cite{boltzman}, we find that today's relic abundances for the species $\psi$ will be given by the expression:
\be\label{relic}
\Omega_{\psi,q} h_0^2= (\Omega_\psi h_0^2)_{no-source}\times(\frac{\tilde{g_{eff}}(\chi_f)}{g_{eff}(\chi_f)})^\frac{1}{2}\times(1+\int\limits_{\chi_0}^{\chi_f}\frac{\Gamma H^{-1}}{\psi(\chi)}d\chi)\times f_{correcrion}(\chi_f)
\ee
\nin where the first term stands for the standard result:
\be
(\Omega_\psi h_0^2)_{no-source}=\frac{1.066\times 10^9 GeV^{-1}}{m_{pl}\sqrt{g_{eff}}J}
\ee
\nin with $J=\int\limits_{\chi_0}^{\chi_f}\left\langle \sigma_A\left|\upsilon\right|\right\rangle d\chi$ and we have defined as in \cite{boltzman},
\be
\psi(\chi)\equiv \chi e^{-\int\limits_{\chi_0}^{\chi}\frac{\Gamma H^{-1}}{\chi}d\chi}
\ee

The last term is the correction braught by the non-extensitivity and has the form:
\be
f_{correction}=1+(\frac{1}{2\tilde{g_{eff}}(\chi_f)}-\frac{1}{g_{eff}(\chi_f)})(g_{eff,q}-g_{eff})-\frac{\left\langle \sigma_A\left|\upsilon\right|\right\rangle (\chi_{f,q}-\chi_f)}{J}+\frac{\chi_{f,q}-\chi_f}{\psi(\chi_f)}(1+\int\limits_{\chi_0}^{\chi}\frac{\Gamma H^{-1}}{\psi(\chi)}d\chi)
\ee
\nin whereas the second and the third term in (\ref{relic}) represent the corrections due to the off-shell/dilaton terms estimated by the authors in \cite{boltzman}.

To do a rough estimation: the modification braught by the SSC term (third term in (\ref{relic})) is of the order of $\frac{\chi_0}{\chi_f}\approx 10^{-13}$ if matter contribution is considered to be negligeable and we also take the off-shell terms to be very small (see\cite{boltzman}) whereas in the last term of (\ref{relic}) the biggest contribution comes with a factor ($q-1$) which in general is greater than $10^{-13}$. Therefore the effect of non-extensitivity on today's relic abundances can be significant for appropriate choices of q and as such can be used to put constraints on the model by comparison with the corresponding phenomenology.

\section{D-particles foam model}
In this part we briefly present a physical model that as we show, can give rise to cosmological results similar (not necessarily identical) to those of Tsallis statistics.The basic idea for this model, is that the spacetime is viewed as a "fuzzy" entity, revealing this way its dynamical aspect imposed by a general relativistic point of view .The origin of this "fuzziness" within the context of quantum gravity, is found in the interactions of point-like stringy defects of spacetime (D-particles) with closed strings.Then as a result of the momentum transfer during the interaction, there is a D-particles recoil and this can happen in a stochastic way, meaning that any fraction of the momentum of the string could be transferred to the D-particles. Due to this process, the spacetime behaves as a randomly fluctuating environment and this is expressed through microscopic fluctuations of the metric. Therefore if one considers the case of flat spacetime, a generic metric would have the form:

\be\label{metric}
\\g_{\mu\nu}=
\left(
\begin{array}{cccc}
       -1    &  k_1r_1  &  k_2r_2  & k_3r_3 \\
      k_1r_1 &   1      &    0     &    0   \\
      k_2r_2 &   0      &    1     &    0   \\
      k_3r_3 &   0      &    0     &    1
 \end{array}
 \right)
 \ee
 \nin where in this model we have chosen the off-diagonal metric components $g^{0i}$ to depend on the momentum of the particle as well as on a set of statistical parameters $(r_1,r_2,r_3)$ that describe the amount of momentum that is transferred in each collision to each spatial direction.
 It is important to note that these parameters are assumed to follow a gaussian (normal) distribution with a standard deviation $\sigma_i$ that is not constant but is distributed with a $\chi^2$ distribution , i.e. its average value will be:
 \be
 <\sigma_i>=\int\limits_{0}^{\infty}{\sigma_i f(\sigma_i) d\sigma_i}=\sigma_{i0}
 \ee
 \nin with the probability density given by:
 \be
 f(\sigma_i)=\frac{1}{\Gamma(\frac{n}{2})}(\frac{n}{2\sigma_{i0}})^\frac{n}{2}\sigma_i ^{\frac{n}{2}-1}\exp(\frac{-n\sigma_i}{2\sigma_{i0}})
 \ee
 
 In order to derive the statistical description for particle systems in our model, we need to obtain an expression for the particles energy in a quantum level approach. Let us take the massive Klein-Gordon equation in a general gravitational background:
 \be
 [g^{\mu\nu}\partial_\mu\partial_\nu-m^2]\Phi=0
 \ee
 
 If we expand this form and we also assume plane wave solutions, we obtain the following expression for the particle energy:
 \be\label{energy}
 \omega_{1,2}=\frac{2g^{0i}k_i\pm\sqrt{4(g^{0i}k_i)^2-4g^{00}(g^{ii}(k_i)^2+m^2)}}{2g^{00}}
 \ee
 \nin where we have thrown away all cross terms, that is terms of the form $k_ik_j$ (see also \cite{bernabeu}).
 
 Subsequently, we find what (\ref{energy}) gives for the case of the metric (\ref{metric}). The only approximation we make is to drop again all cross terms.
 In order also to simplify our expression we make the substitution: $E^2=k_1^2+k_2^2+k_3^2+m^2$ and this in fact would correspond to the particle energy in the case of a Minkowski spacetime. But in our case ((\ref{metric})) this is just a NOTATION and nothing more since the actual particle energy is given by the expression (\ref{energy}). 
 Now with the calculated energy in hand and by using the standard definition of statistical mechanics: $<n>=\frac{1}{\exp(\beta(\omega_r-\mu))+\xi}$, we can obtain the distribution functions for fermions and bosons in our model:
\bea\label{dis}
&&<n>=\frac{1}{\exp(\beta(E-\mu))+\xi}\nn
&& -\frac{\exp(\beta(E-\mu))}{(\exp(\beta(E-\mu))+\xi)^2}\frac{\Gamma(\frac{n}{2}+2)}{\Gamma(\frac{n}{2})}\frac{2}{n^2}\nn
&&[(\beta^2+\frac{\beta}{E})(\sigma_{01}^2k_1^4+\sigma_{02}^2k_2^4+\sigma_{03}^2k_3^4)+\frac{\beta}{E}m^2(\sigma_{01}^2k_1^2+\sigma_{02}^2k_2^2+\sigma_{03}^2k_3^2)]
\eea
\nin where $\xi=+1$ applies to fermions and $\xi=-1$ applies to bosons.
By recalling that $\sigma_{0i}$ has dimensions of inverse temperature (see \ref{metric}) it is convenient to set: $\sigma_{0i}=\frac{\hat{\sigma_{0i}}}{T}$, where $\hat{\sigma_{0i}}$ is now dimensionless.Under this substitution, (\ref{dis}) becomes:
\bea\label{difun}
&&<n>=\frac{1}{\exp(\beta(E-\mu))+\xi} \nn
&&-\frac{\exp(\beta(E-\mu))}{(\exp(\beta(E-\mu))+\xi)^2}\frac{\Gamma(\frac{n}{2}+2)}{\Gamma(\frac{n}{2})}\frac{2}{n^2}\nn
&&[(\beta^4+\frac{\beta^3}{E})(\hat{\sigma_{01}}^2k_1^4+\hat{\sigma_{02}}^2k_2^4+\hat{\sigma_{03}}^2k_3^4)+\frac{\beta^3}{E}m^2(\hat{\sigma_{01}}^2k_1^2+\hat{\sigma_{02}}^2k_2^2+\hat{\sigma_{03}}^2k_3^2)]
\eea

This is our expression for the distribution functions from which one can reproduce all number and energy denisties for bosons and fermions in both the relativistic and non-relativistic limit.

\section{Conclusions}
We saw that when Tsallis statistics was applied to one case of SSC we obtained quite interesting cosmoligical effects. At first we obtained an "exotic" scaling for the radiation energy density since our analysis referred to a r.d.e, which seemed to be a rather generic effect. In fact if the same analysis had been performed for a m.d.e, fractal scaling for "dust" matter should occur as well. The second interesting effect has been the modifications to the relic abundances of CDM candidates (neutralinos) which seem to dominate on the effects due to off-shell/dilaton terms presented in \cite{boltzman}. These modifications should in fact be constrained by the relevant phenomenology but this is a part of another work. 
In the last part of our analysis,we saw that by using this modified metric (\ref{metric}) we obtained a distribution function expression ((\ref{difun}) that is very similar to the result obtained in \cite{pessah}, where the modified distribution functions were derived by introducing a non-additive parameter q to the standard Boltzman-Gibbs entropy (Tsallis entropy).The smallness of the deviation from the standard case (1st term in (\ref{difun})) was guaranteed in our case by the smalness of the statistical parameters $\hat{\sigma_{0i}}$ whereas in the Tsallis statistics case this came from requiring small departure of q from the value 1. Therefore,we have proven that we managed to get a physical result very similar to the result of  Tsallis statistics by simply starting from a different background spacetime physically stemming from stringy defects (D-particles) recoil.The same analogy holds for all the other quantities defined in\cite{pessah} since we are performing our analysis with equivalent distribution functions.However those results will appear in a future work together with the effects on dark matter relic abundances since in this part of our analysis our main goal was restricted to demonstrating the analogy between the physical effects of Tsallis statistics and the D-particles foam model.
 
\section*{Acknowledgements}
I would like to thank N.E. Mavromatos and S.Sarkar for their very helpful suggestions throughout all this work.

\end{document}